\documentclass[a4paper,twocolumn,showpacs,preprintnumbers,amsmath,amssymb]{revtex4}
\usepackage{graphicx}
\usepackage{dcolumn}
\usepackage{bm}
\usepackage{geometry}
\usepackage{amsmath}
\usepackage{literat}
\newcommand{\ve}{\varepsilon}

\begin{document}
\title{Self-organization processes in laser system with nonlinear absorber and
external force influence}
\author{E.D.~Belokolos, V.O.~Kharchenko}\email{vasiliy@imag.kiev.ua}
\affiliation{Institute of magnetism, National Academy of Science of Ukraine,
03142, Kiev, Ukraine}
\date{\today}
\begin{abstract}
We discuss mechanisms of self-organization processes in two-level solid-state
class-B laser system. The model is considered under assumptions of influence of
nonlinear absorber and external force, separately. It was found that
self-organization occurs through the Hopf bifurcation and results to a stable
pulse radiation. Analysis is performed according to the Floquet exponent
investigation. It was found that influence of the nonlinear absorber extends
the domain of control parameters that manage a stable periodic radiation
processes. An external force suppresses self-organization processes. A combined
influence of both external force and nonlinear absorber results to more
complicated picture of self-organization with two reentrant Hopf bifurcations.
\end{abstract}

 \maketitle

\section{Introduction}

The most intriguing phenomena in systems with nonlinear dynamics is a
transition to the regime with dissipative structures formation. A related
problem of such effects investigation in systems with large numbers of freedom
degrees attracts an increasing attention in last three decades. Due to
self-organization effects a number of freedom degrees is reduced and
description of the system dynamics can be performed in terms of macroscopic
variables. A typical picture is realized in laser systems where description is
provided with a help of amplitudes of electric field (or intensity of the
radiation), polarization and population inversion \cite{Haken}.

Laser theory shows that corresponding dissipative structures define formation
of pulse or modulated signals in homogeneous systems or spirals in spatially
extended ones \cite{zigzag}. In practice a formation of stable periodic
radiation can be induced by introducing an additional medium with nonlinear
properties which are realized as an absorber or modulator. Such type of lasing
is known as passive one. Usually, such a kind of medium leads to nonlinear
dependence of the relaxation time of the electric field amplitude
\cite{Khanin1,Khanin2} or nonlinear dependence of refractive exponent
\cite{Hercher,Hercher2,physrevlet76,b716}, composite material can be used to
introduce different type of such nonlinearity
\cite{sdarticle4,citation2,citation3,citation4,citation5,citation6}. A coherent
dynamics of two-level laser systems in the presence of dispersive and
absorptive effects was observed theoretically and experimentally
\cite{exp2003,numeric2002}. Statistical properties of self-organization effects
of such type systems was discussed in \cite{condmat1}. Regimes of optical
parametric oscillation in a semiconductor microcavity are studied in
\cite{condmat2}. It was found that stationary behaviour of polarization can be
described by the formalism of non-equilibrium transitions, where bistability is
observed (see \cite{physrevlet76,PhysRevA78}). It was shown that an oscillating
lasing is realized inside a bounded domain of the system parameters. Another
(active) way to initiate a coherent lasing is an introducing an external
influence on nonlinear processes in the cavity \cite{kachmarek}. An actual
problem in laser physics is to find possible mechanisms and to set a range of
control parameters that manage properties of stable periodic radiation (see for
instance \cite{Khanin1,Khanin2} and citations therein). Despite this problem is
still opened in deterministic (regular) systems a lot of attention is paid to
find coherent regimes under influence of stochastic sources
\cite{Gardiner86,Gardiner2000,Risken84,Horstshemke}.

In this Paper, we are aimed to investigate the dynamics of the solid-state
class-B laser systems which are simple in realization and are wide used in
physical applications. We consider deterministic models only. According to the
theoretical approach, based on Floquet analysis, we will explore in what a
manner a nonlinear medium can induce a stable periodic radiation. It will be
shown that varying in a saturation amplitude of electric field and absorption
coefficient one can arrive at stable and unstable dissipative structures.
Properties of self-organization process induced by an external force influence
will be considered. At last, a combined influence of both nonlinear medium and
external force on the system dynamics will be described.

The paper is organized in the following manner. In Section II we present a
model of our system where we introduce theoretical constructions to model an
influence of both an absorber and external force. Section III is devoted to
development of the analytical approach to study process of dissipative
structures formation. In Section IV we apply the derived formalism to
investigate properties of stable periodic radiation in the presence of the
absorber, external force and its combined affect. Main results and perspectives
are collected in the Conclusion (Section V).

\section{Model}

Considering a prototype model for a two-level laser system, one deals with
dimensionless variables such as: an electric field amplitude $E$, polarization
$P$ and $S$ to be a population inversion. A standard technique usage
\cite{Khanin1} allows to reproduce evolution equations for these three
macroscopic freedom degrees from both the Maxwell-type equation for
electro-magnetic field and density matrix evolution equation. It leads to the
system of Maxwell-Bloch type that is reduced to the Lorenz-Haken model in the
form
\begin{equation}
 \left\{
  \begin{array}{l}
   \varkappa^{-1}\dot{E}=-E+P,\\
   \gamma_\bot^{-1}\dot{P}=-P+ES,\\
   \gamma_\|^{-1}\dot{S}=(S_e-S)-EP.
  \end{array}
 \right.
 \label{lor}
\end{equation}
For the single mode laser system a relaxation of electric field amplitude $E$
is addressed to losses in a bulk of the medium and characterized by the
velocity $\varkappa=1/2\tau_c$, where $\tau_c$ is a life-time of a photon in a
cavity. $\gamma_\bot $ is a relaxation velocity of nondiagonal elements of
density matrix which is related to the half-width of a spectral line. The
relaxation scale for the population inversion is determined by the velocity
$\gamma_\|$ defined by both transition probability between two energy levels
and a corresponding frequency. $S_e$ controls the pump intensity, as usual. The
model (\ref{lor}) shows a linear combination of the amplitude $E$ and
polarization $P$, despite the evolution of both $P$ and the pump intensity $S$
are nonlinear. It is principally that the positive feedback of $E$ and $S$
leads to instability in the polarization that induces a self-organization.
According to the Le-Shatelier principle such positive feedback is compensated
through negative one in third equation (the last term).

To make an analysis we pass to dimensionless variables $\tau\equiv t\varkappa$,
$\sigma\equiv\varkappa/\gamma_\bot$ and $\ve\equiv\varkappa/\gamma_\|$. Hence,
the system (\ref{lor}) takes the form
\begin{equation}
\left\{
\begin{array}{l}
 \dot{ E}=- E+P,\\
 \sigma\dot{P}=-P+ E S,\\
 \ve\dot{S}=(S_e-S)- E P.
\end{array}
\right. \label{lor3eq}
\end{equation}
Assuming different combinations between relaxation scales $\sigma$ and $\ve$,
one can describe three possible classes of laser systems. At $\ve,\sigma\ll 1$
we arrive at the laser models of class-A (organic dye lasers) with
one-dimensional phase space, where systems states are represented by fixed
points only. Here self-organization effects are described by a formalism of
non-equilibrium phase transitions. Class-B (solid-state lasers) is
characterized by a condition $\sigma\ll 1$. Here phase space is two-dimensional
and transition processes are of oscillation type and hence self-organization
processes result in dissipative structures formation. For the class-C
(molecular gas lasers) we set $\sigma,\epsilon\sim 1$ and in three-dimensional
phase space a strange attractor can be realized. At last, the class-D (beam
masers) is characterized by condition $\sigma,\epsilon\gg 1$. In this Paper we
consider the class-B only, where the polarization $P$ is assumed to be a
microscopic quantity and should be treated as fast variable which follows the
electric field amplitude $E$ evolution. Such situation is realized in single
mode solid laser systems with low-doped crystals ($Al_2O_3:Cr^{3+}$) and
glasses (soda-lime glass), some gas lasers ($CO_2$), fiber and semi-conductor
lasers \cite{Khanin1,Khanin2,PRA2002}.

Assuming conditions $\gamma_\bot\gg\varkappa,\gamma_\|$, one can use the
adiabatic elimination procedure which yields the relation $P=ES$. As a result,
instead of the system (\ref{lor3eq}) we obtain a two-component model in the
form
\begin{equation}
 \left\{
  \begin{array}{l}
   \dot{E}=-E(1-S),\\
   \dot{S}=\ve^{-1}\left[S_e-S(1+E^2)\right].
  \end{array}
 \right.\label{lor2eq}
\end{equation}

The model (\ref{lor2eq}) can not show the stable oscillating regime of the
electric field $E$, itself. It was shown experimentally and theoretically
\cite{Hercher,Khanin1,Khanin2} that stable oscillations can be realized if an
additional nonlinear medium is introduced into the cavity. The first way to get
the periodic lasing is to use a passive modulating medium (nonlinear material)
to absorb a weak radiation and transmit signal with large amplitude. Such a
type of absorbers is realized in practice as phthalocyanine fluid in
Fabry--Perot cavities \cite{Optics2002}, gases $SF_6$, $BaCl_3$ and $CO_2$
\cite{PismavGETF,GETF71}. To describe action of the absorber it was proposed to
introduce a nonlinear damping into evolution equation for the electric field
\cite{Haken80}
\begin{equation}\label{f_k}
f_\kappa=-\frac{\kappa E}{1+E^2/E_s^2},
\end{equation}
here $E_s$ is the saturation amplitude. The second way is to use an additional
medium with nonlinear refractive exponent $n=n(E)$
\cite{Hercher,Hercher2,sdarticle4,physrevlet76}. Such type of modulator can be
used to increase the Q-factor of laser. We will model action of such an
effective medium by the external force $f_e(E)$ assumed in the form
\begin{equation}\label{V(E)}
f_e=-A-CE^2,
\end{equation}
that correspond to action of a bare potential $V=AE+CE^3/3$, where coefficients
$A$, $C$ controls photon processes in the modulator. We use an general
construction (\ref{V(E)}) in order to investigate an influence of parameters
$A$ and $C$ on lasing. In physical applications one can associate $-A$ as
incident field amplitude, $C$ can control nonlinear properties of the
refractive index $n(E)$. One of the simplest situations is considered in
\cite{PhysRevA78}, where only a case of $A<0$ was investigated.

Combining all above suppositions into the one model for a single-mode laser
system, we will get the generalized system of nonlinear equations type of
\begin{equation}\label{lor2gen}
 \left\{
  \begin{array}{l}
   \dot{E}=-E(1-S)+f_e(E)+f_\kappa(E),\\
   \dot{S}=\ve^{-1}\left[S_e-S(1+E^2)\right].
  \end{array}
 \right.
\end{equation}
Using two type of additional medium in the cavity, one can expect that some
combinations of parameters for both modulator and absorber should exist to
provide the stable periodic radiation of the laser.

\section{Main equations}

To find mechanisms which takes care of the stable dissipative structures
formation we will use the standard procedure to analyze conditions where
bifurcation into limit cycle occurs \cite{hassard}. To this end we rewrite the
system (\ref{lor2gen}) in a most general form
\begin{equation}
\left\{
\begin{array}{l}
 \dot{E}=f^{(1)}(E,S),\\
 \ve\dot{S}=f^{(2)}(E,S),
\end{array}
\right. \label{sec4eq1}
\end{equation}
where effective forces are as follows:
\begin{equation}
\begin{split}
 &f^{(1)}(E,S)\equiv-\left[A+CE^2\right]-E-\frac{\kappa E}{1+E^2/E_s^2}+ES,\\
 &f^{(2)}(E,S)\equiv\ve^{-1}\left[S_e-S(1+E^2)\right],
\end{split}
\label{sec4eq2}
\end{equation}\label{potential}
here constructions (\ref{f_k}), (\ref{V(E)}) are used.

We deal with a problem of nonlinear dynamics and present a behaviour of the
system in the phase plane $(E,S)$. Firstly, we consider steady states $E_0$ and
$S_0$, defined as coordinates of fixed points in the phase plane. Setting $\dot
E=0$ and $\dot S=0$, one can find steady states as solutions of stationary
equations
\begin{equation}
\begin{split}
 &E_0\left(\frac{S_e}{1+E_0^2}-\frac{\kappa E_s^2}{E_0^2+E_s^2}-2CE_0-1\right)=A,\\
 &S_0=S_e(1+E_0^2)^{-1}.
\end{split}
 \label{sec4eq3}
\end{equation}
A behaviour of phase trajectories in the vicinity of these fixed points can be
analyzed with a help of the Lyapunov exponents approach. Here time dependent
solutions of above system are assumed to be in the form $ E\propto e^{\Lambda
t},\quad \Lambda=\lambda+i\omega$, where $\lambda$ controls the stability of
the phase trajectories, $\omega$ determines pulse frequency of the signal.
Magnitudes for real and imaginary parts of $\Lambda$ are calculated according
to the Jacobi matrix elements
\begin{equation}
M_{ij}\equiv\left(\frac{\partial{f^{(i)}}}{\partial{x_j}}\right)_{x_j=x_{j0}};
\quad x_j\equiv\{ E,S\},\quad i,j=1, \label{sec4eq4}
\end{equation}
where subscript 0 relates to steady states. Inserting (\ref{sec4eq2}) into
definition (\ref{sec4eq4}), we get matrix elements
\begin{eqnarray}
  &&M_{11}=-M_0+S_0,\\ \nonumber
  &&M_0=1+2CE_0+\kappa\frac{1-E_0^2/E_s^2}{(1+E_0^2/E_s^2)^2},\\
  &&M_{12}= E_0;\ M_{21}=-2\ve^{-1}S_0 E_0;\\
  &&M_{22}=-\ve^{-1}(1+ E_0^2).\nonumber
\label{sec4eq5}
\end{eqnarray}
Then, an equation for eigenvalues and eigenvectors
\begin{equation}
\sum_j M_{ij}V_j=\Lambda V_i
\label{sec4eq6}
\end{equation}
gives expressions for $\lambda$ and $\omega_0$ as follows:
\begin{equation}
\begin{split}
 &\lambda =\frac{1}{2}\left[(S_0-M_0)-\ve^{-1}\left(1+ E_0^2\right)\right],\\
 &\omega_0=\frac{1}{2}\sqrt{8\ve^{-1}S_0 E_0^2-\left[(S_0-M_0)+\ve^{-1}(1+ E_0^2)\right]^2}.
\end{split}
\label{sec4eq7}
\end{equation}
If the real part of the Lyapunov exponent $\lambda = 0$ then a fixed point
$(E_0,S_0)$ is addressed to a center of a limit cycle. It leads to relation
\begin{equation}
\ve(S_0-M_0)\geq 1+ E_0^2; \label{sec4eq8}
\end{equation}
and yields a condition for the frequency of oscillations
\begin{equation}
8\ve S_0 E_0^2\geq \left[\ve (S_0-M_0)+(1+
E_0^2)\right]^2.\label{sec4eq9}
\end{equation}

To investigate a stability of such a limit cycle we analyze a behaviour of
trajectories in the vicinity of the fixed point $(E_0,S_0)$. To this end we
rewrite motion equations (\ref{sec4eq1}) where variables $E$ and $S$ are count
off from stationary magnitudes $E_0,S_0$. To do this one can use following
transformation
\begin{equation}
\vec{X}=\vec{X}_0+\hat P\cdot\vec{\delta}, \label{sec4eq10}
\end{equation}
where notations for pseudovectors are used:
\begin{equation}
\vec{X}\equiv\left(
\begin{array}{l}
 E\\S
\end{array}
\right),\quad \vec{\delta}\equiv\left(
\begin{array}{l}
 E- E_0\\S-S_0
\end{array}
\right).
 \label{sec4eq11}
\end{equation}
The corresponding transformation matrix $\hat P$ is obtained with a help of
eigenvector $\vec{V}$ components, i.e.:
\begin{equation}
P\equiv\left(
\begin{array}{l}
\Re V_1\quad -\Im V_1\\ \Re V_2\quad -\Im V_2
\end{array}
\right),\quad \vec{V}\equiv\left(
\begin{array}{l}
V_1\\V_2
\end{array}
\right). \label{sec4eq12}
\end{equation}

Assuming $V_1\equiv 1$, for the second component $V_2$ from Eq.(\ref{sec4eq6})
one gets
\begin{equation}
\begin{split}
V_2&=\frac{(M_0-S_0)+{\rm i}\omega_c }{ E_0},\\
\omega_c&\equiv\left.\omega_0\right|_{\lambda=0}=\ve^{-1}(1+
E_0^2)\left[\frac{2S_e E_0^2}{(1+ E_0^2)^3}\ve-1\right]^{1/2}.
\end{split}\label{sec4eq14}
\end{equation}
Hence, the transformation matrix (\ref{sec4eq12}) takes the form
\begin{equation}
P=\left(
\begin{array}{l}
1\qquad\qquad\quad\quad\quad 0\\ (M_0-S_0)/ E_0 \quad -\omega_c /
E_0
\end{array}
\right). \label{sec4eq15}
\end{equation}
It leads to evolution equations for deviations written in a vector form
\begin{equation}
\dot{\vec{\delta}}=\vec{F},\qquad \vec{F}\equiv P^{-1}\vec{f}. \label{sec4eq16}
\end{equation}
Here a pseudovector of the canonical force
\begin{equation}
\vec{F}=\left(
\begin{array}{l}
F^{(1)}\\F^{(2)}
\end{array}
\right)\equiv\left(
\begin{array}{l}
f^{(1)}-f^{(1)}_0\\f^{(2)}-f^{(2)}_0
\end{array}
\right), \label{sec4eq17}
\end{equation}
satisfies conditions \cite{hassard,Poincare,Andronov,Leontovich}
\begin{equation}
\frac{\partial\vec{F}}{\partial\vec{\delta}}= \left(
\begin{array}{l}
0\quad -\omega_c \\\omega_c \qquad 0
\end{array}
\right), \label{sec4eq18}
\end{equation}
and has following components:
\begin{equation}
F^{(1)}=f^{(1)},\quad F^{(2)}=\alpha f^{(1)}-\beta\ve
f^{(2)}; \label{sec4eq19}
\end{equation}
\begin{equation}
\alpha \equiv\frac{M_0-S_0}{\omega_c },\qquad \beta \equiv\frac{
E_0}{\ve \omega_c }. \label{sec4eq20}
\end{equation}

Above procedure allows to find the stability of the manifold formed by the
fixed point $(E_0, S_0)$. Using the standard technique \cite{hassard}, one can
say that the limit cycle is stable only if a real part of the Floquet exponent
\begin{equation}
\Phi=\frac{\rm i}{2{\omega_0}} \left(
g_{11}g_{20}-2|g_{11}|^2-\frac{1}{3}|g_{02}|^2\right)
+\frac{1}{2}g_{21}, \label{sec4eq21}
\end{equation}
is negative in a bifurcation point. Structure constants in the definition
(\ref{sec4eq21}) are described by derivatives with respect to $E$ and $S$,
denoted with subscripts:
\begin{equation}
g_{11}=\frac{1}{4}\left[\left(F^{(1)}_{EE}+F^{(1)}_{SS}\right)
+{\rm i}\left(F^{(2)}_{EE}+F^{(2)}_{SS}\right)\right], \label{sec4eq22}
\end{equation}
\begin{equation}
\begin{split}
\left(
\begin{array}{l}
g_{02}\\g_{20}
\end{array}
\right)=\frac{1}{4}&\left[\left(F^{(1)}_{EE}-F^{(1)}_{SS} \mp
2F^{(2)}_{ES}\right)+\right.\\ &\left. {\rm i}\left(F^{(2)}_{EE}-F^{(2)}_{SS} \pm 2F^{(1)}_{ES}\right)\right],
\end{split}
\label{sec4eq23}
\end{equation}
\begin{equation}
\begin{split}
g_{21}=&\frac{1}{8} \left\{\left[\left(F^{(1)}_{EEE}+F^{(1)}_{ESS}\right)+
\left(F^{(2)}_{EES}+F^{(2)}_{SSS}\right)\right]+\right.\\
 &\left.{\rm
i}\left[\left(F^{(2)}_{EEE}+F^{(2)}_{ESS}\right)-
\left(F^{(1)}_{EES}+F^{(1)}_{SSS}\right)\right]\right\}.
\end{split}\label{sec4eq24}
\end{equation}
Using some algebra, the stability condition for the limit cycle can be written
as follows
\begin{equation}
\begin{split}
 &2\alpha(\psi_\kappa-C)^2+\alpha\beta\ve S_0(1+2\beta\ve E_0)+\omega_c(\phi_\kappa+\beta\ve)\leq\\
 &(C-\psi_\kappa)(\alpha^2-1+2\beta\ve S_0+2\alpha\beta\ve E_0),
\end{split}
 \label{sec4eq28}
\end{equation}
where notations $$\psi_\kappa=-2{\frac {k{{\it E_s}}^{2}E\left (-3\,{{
E_s}}^{2}+{E}^{2} \right )}{\left ({{E_s}}^{2}+{E}^{2}\right )^{3}}},$$
$$\phi_\kappa=6\,{\frac {k{{E_s}}^{2}\left (-6\,{E}^{2}{{E_s}}^{2}+{E}^{4} +{{
E_s}}^{4}\right )}{\left ({{E_s}}^{2}+{E}^{2}\right )^{4} }}$$ are used.

\section{Analysis of Hopf bifurcations}

\subsection{Influence of nonlinear absorber}

\begin{figure}[!h]
\includegraphics[width=70mm]{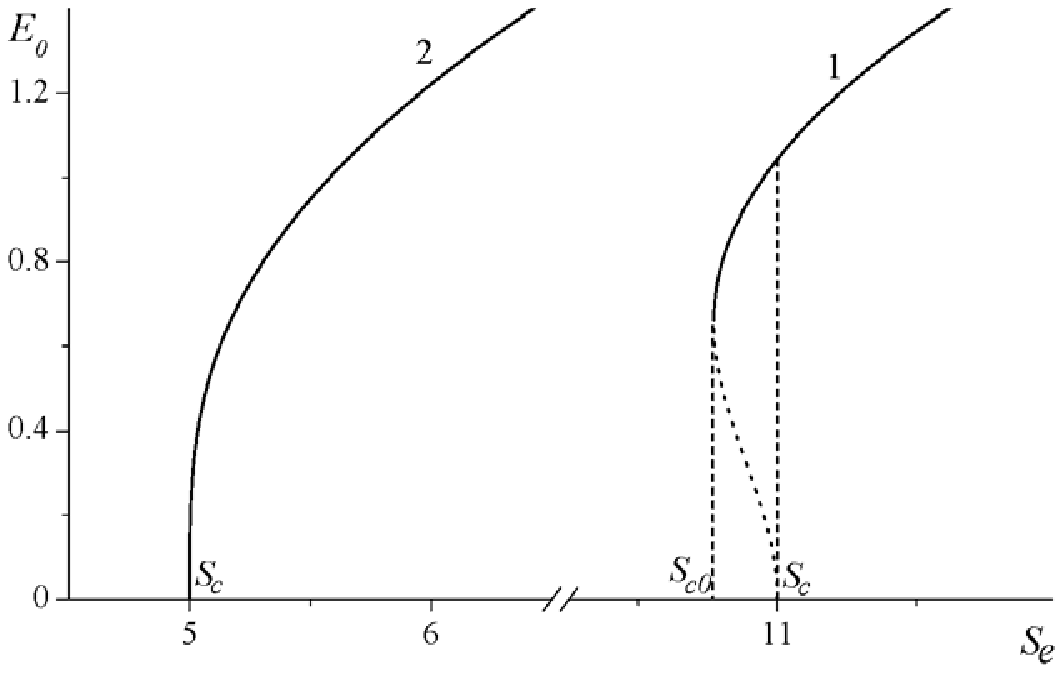}
\centering \caption{Stationary amplitude $E_0$ vs. pump intensity $S_e$ at
different values of absorption coefficient at $E_s=0.9$: curve 1 --
$\kappa=10.0$; curve 2 -- $\kappa=4.0$\label{fig1}}
\end{figure}

To proceed let us consider steady states behaviour under supposition that
action of the absorber is given by expression (\ref{f_k}), $f_e=0$. Setting
$\dot{E}= \dot{S}=0$, one gets stationary values of the electric field
amplitude $E_0$ shown in Fig.\ref{fig1}. A steady states analysis allows to
find that a bistable regime is realized only if $\kappa<\kappa_{min}$, here
$\kappa_{min}=E_s^2/(1-E_s^2)$. In such a case one gets the hysteresis loop in
$E_0(S_e)$ dependence in the domain $[S_{c0}, S_c]$ (curve 1) which disappears
when the threshold $\kappa_{min}$ is crossed, where
\begin{equation}
S_c=1+\kappa,\quad S_{c0}=1+E_s\sqrt{\frac{\kappa}{1-E_s^2}}.
\end{equation}
The behaviour of the amplitude $E_0$ is the same as in the first order phase
transitions where zero value of $E_0$ below $S_{c0}$ corresponds to a
disordered state, values $E_0\ne 0$ (solid line) relate to an ordered state,
whereas intermediate magnitudes of $E_0$ (dotted line) correspond to unstable
state. The critical value for the absorption coefficient is realized only if
the saturation amplitude $E_s<1$. In opposite case one can get the stationary
picture of the second order phase transition where $E_0$ increases
monotonically from 0 if the critical value $S_c$ is crossed (curve 2).

\begin{figure}[!h]
\centering
\includegraphics[width=70mm]{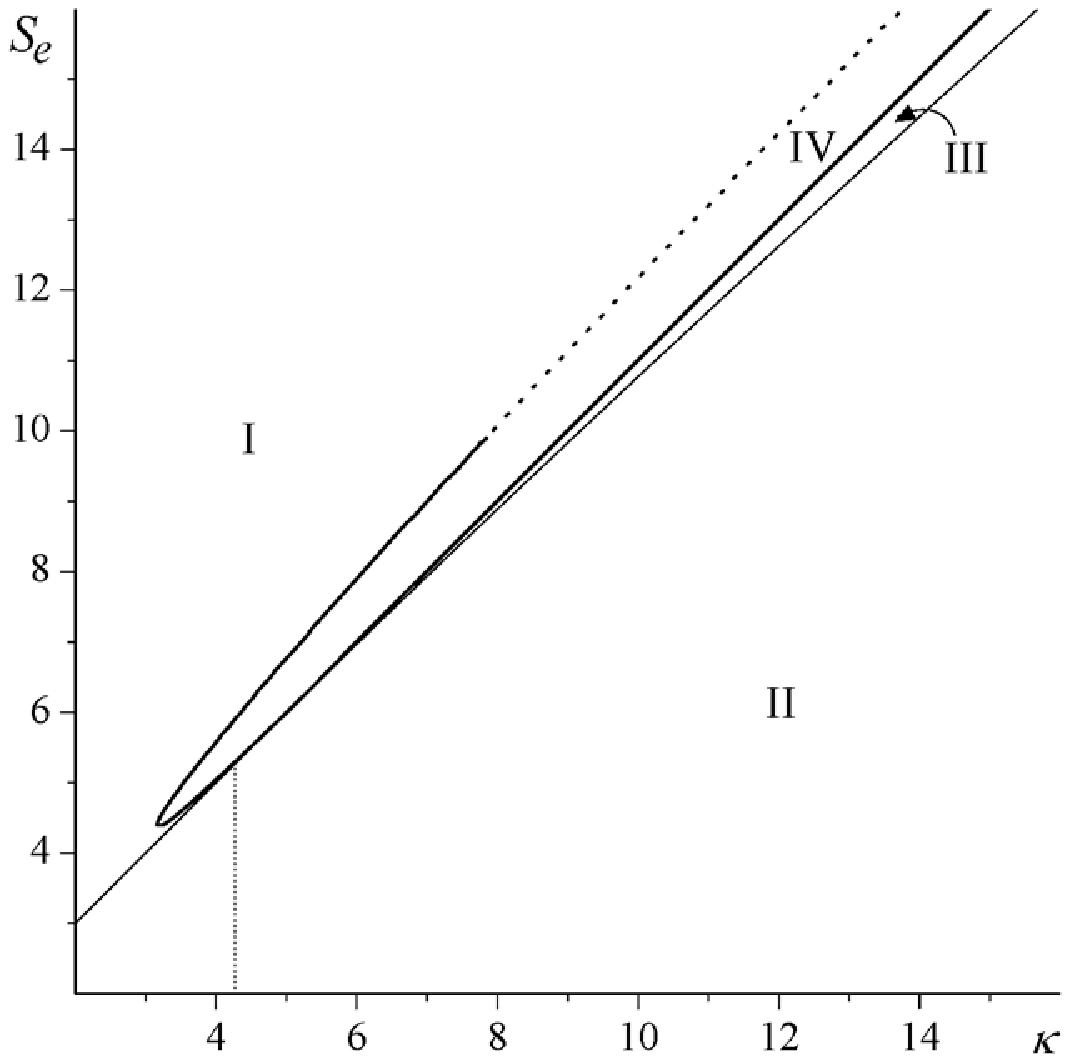}
\caption{Phase diagram in the presence of the nonlinear absorber at
$E_s=0.9$\label{fig2}}
\end{figure}

\begin{figure}[!h]
\centering
 a)\includegraphics[width=35mm]{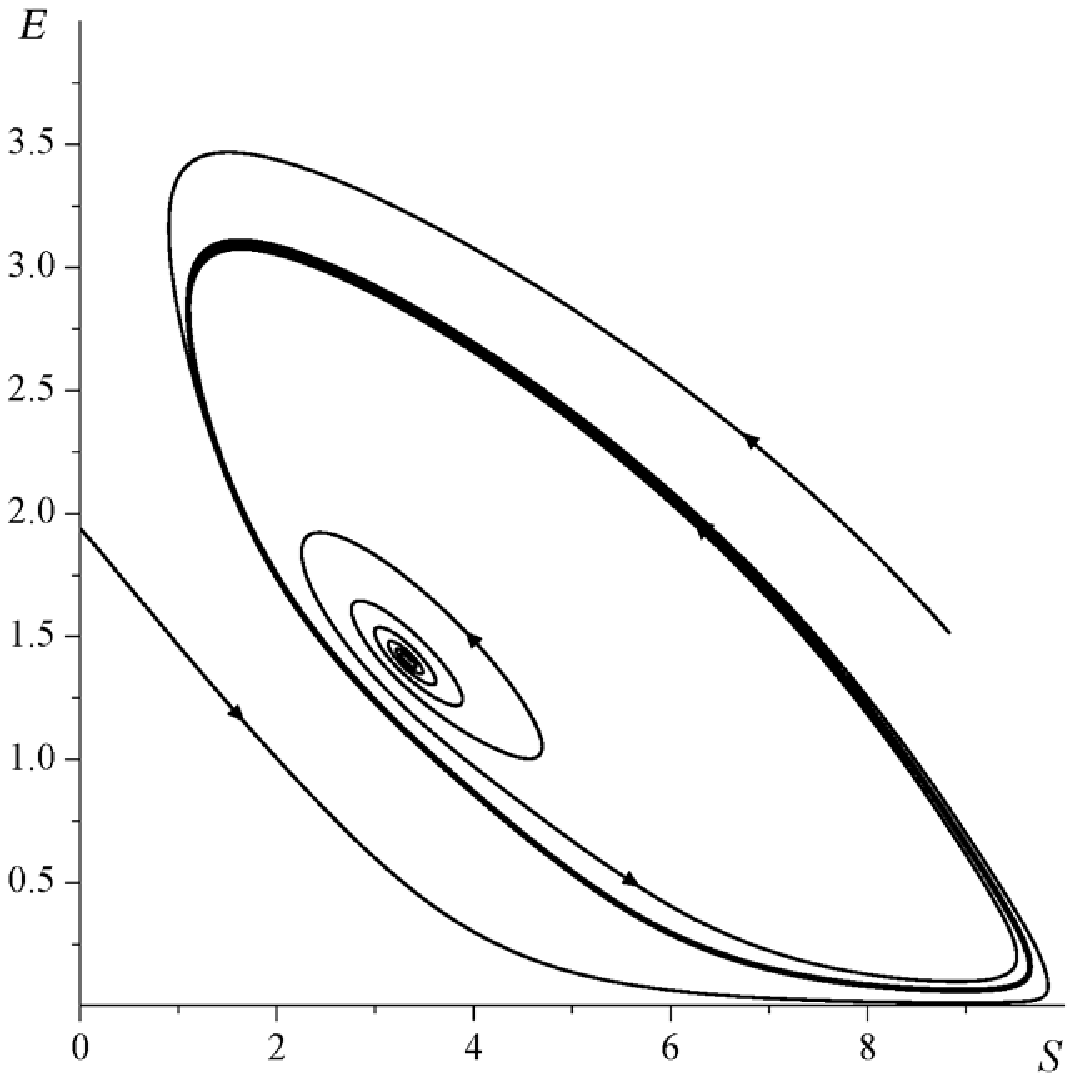}b)\includegraphics[width=35mm]{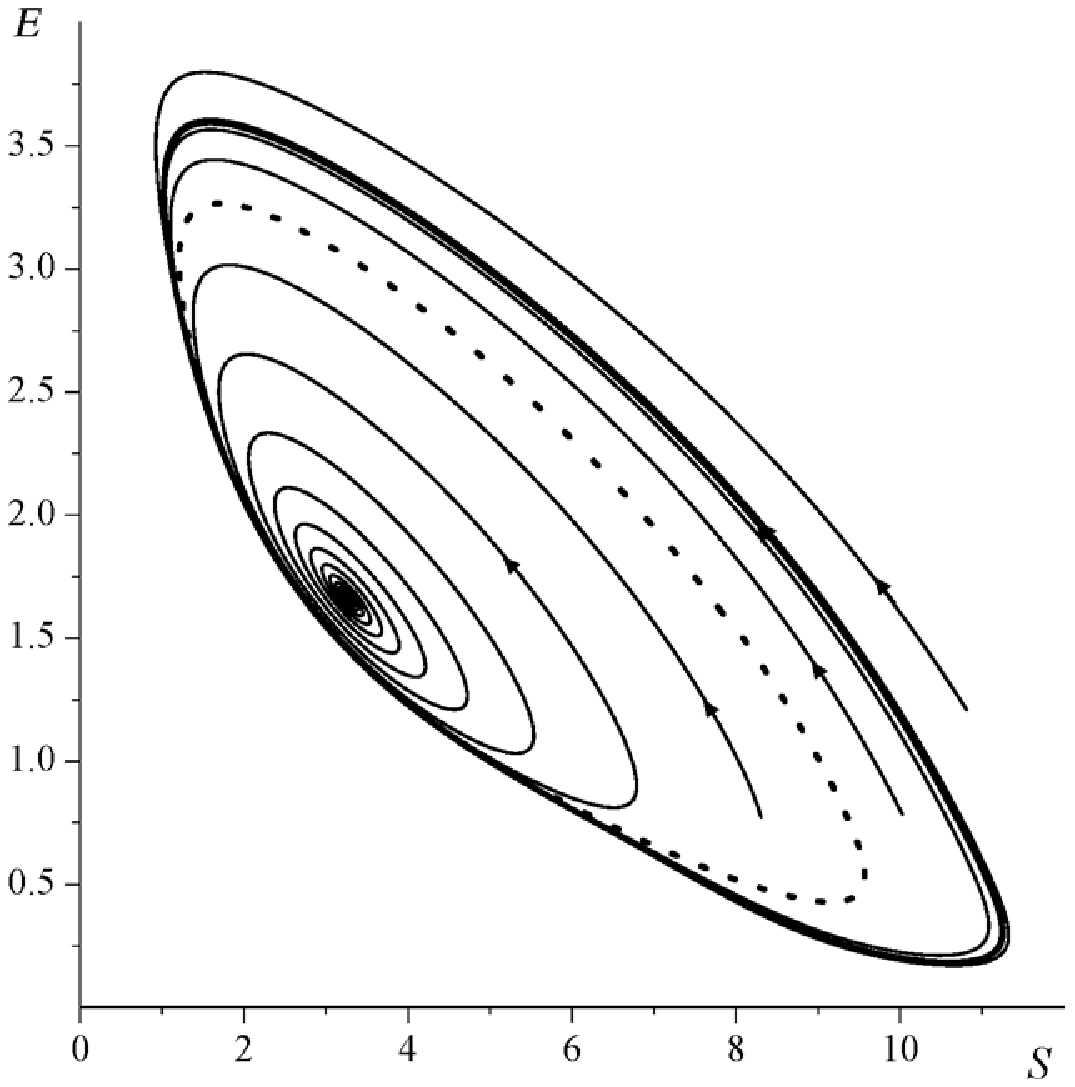}
\caption{Dissipative structures corresponding to domain IV (a) and dotted line
(b) in Fig.\ref{fig2}. Stable and unstable limit cycles: (a) --- $E_s=0.9$,
$\kappa=8.0$, $S_e=9.9$; (b) $E_s=0.9$, $\kappa=10.0$, $S_e=12.3$\label{fig3}}
\end{figure}

\begin{figure}[!h]
\centering
\includegraphics[width=70mm]{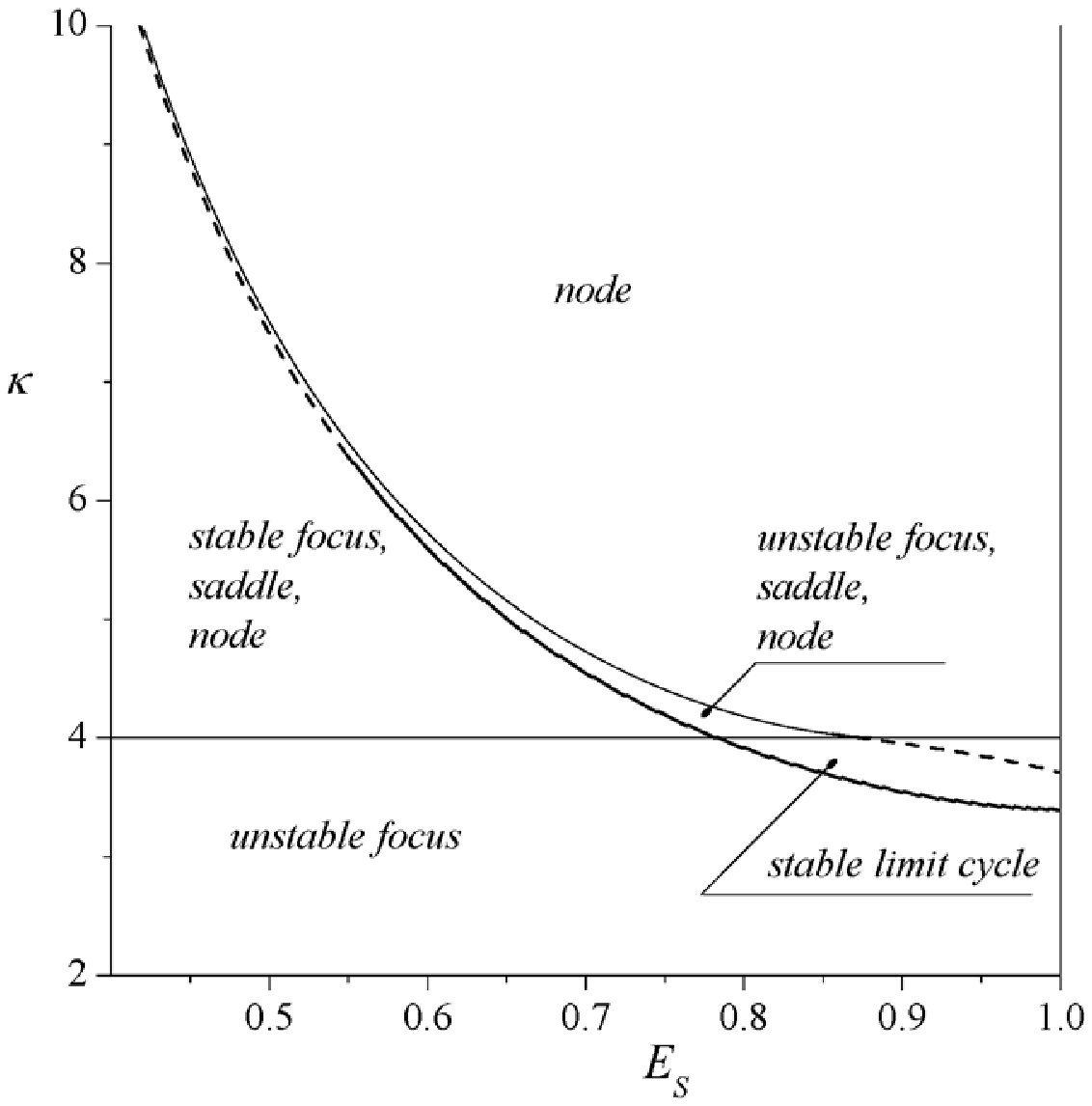}
\caption{Transformation of topology phase plane under influence of the absorber
parameters $\kappa$, $E_s$ at $S_e=5.0$\label{fig4}}
\end{figure}

The analysis of the Floquet exponent allows to find the phase diagram
(Fig.\ref{fig2}), which shows the stable periodic radiation (formation of limit
cycle in the phase plane $(E,S)$).
In Fig.\ref{fig2} the domain I defines configuration of the phase space with
both a stable focus (ordered state) and a saddle point (disordered state); in
the domain II only disordered state is realized (node); the domain III is
characterized by the hysteresis loop, where ordered state corresponds to
unstable focus, unstable state is represented by a saddle, disordered state is
a node. Inside the domain IV the stable limit cycle is formed
(Fig.\ref{fig3}a), which transforms into stable focus, unstable and stable
cycles if dotted line is crossed (Fig.\ref{fig3}b).
An influence of the parameters of the absorber on a topology of phase plane is
shown in Fig.\ref{fig4}.
Here an increase in the absorption coefficient $\kappa$ at small $E_s$ leads to
transformation of unstable focus into a stable one with additional node and
saddle points appearing. At values $\kappa$ and $E_s$, corresponding to the
dashed line, one gets an unstable limit cycle and in the domain bounded by
dashed and solid lines one gets the unstable focus, node and saddle. When the
solid line is crossed the phase portrait is characterized by a single node. An
increase in $\kappa$ at saturation amplitude $E_s\simeq 1$ transforms an
unstable focus into a stable limit cycle, which becomes unstable at values that
correspond to the dashed line. In the domain bounded by the dashed and straight
horizontal lines there is a single unstable focus only. A further increase in
$\kappa$ transforms this focus into a node.

\begin{figure}[!h]
\centering
\includegraphics[width=70mm]{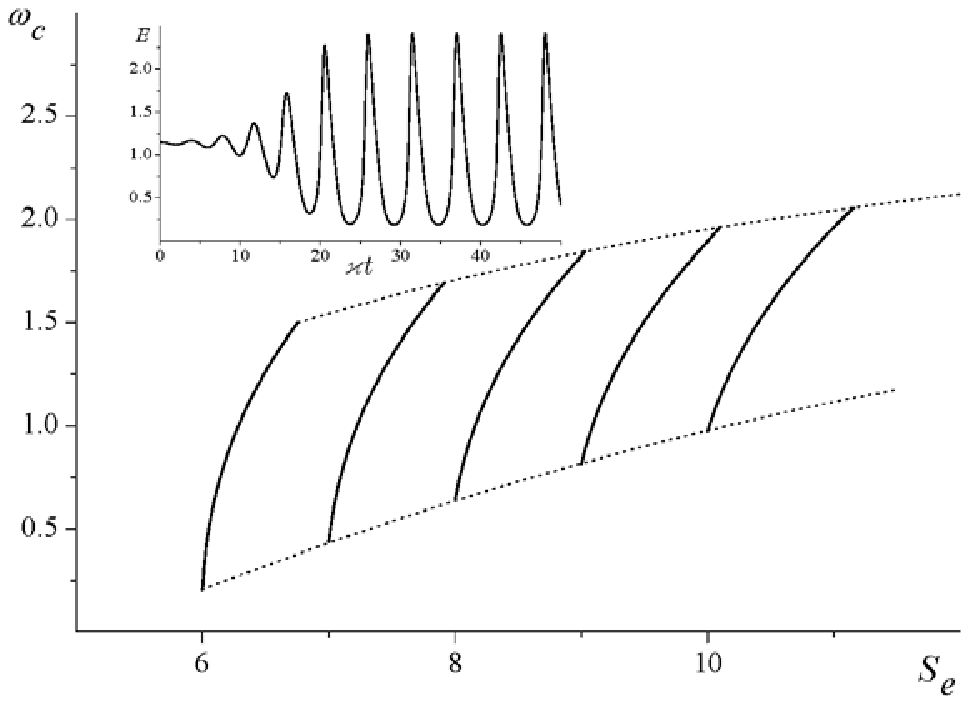}
\caption{Dependence of the stable pulse radiation frequency $\omega_c$ vs. pump
intensity $S_e$ for the system with absorption effect at $E_s=0.9$ and
$\kappa=5$, 6, 7, 8, 9 (from the left to the right). Insertion shows a typical
stable pulse signal\label{fig5}}
\end{figure}

The frequency of pulse radiation regime appears at non zero value, that
correspond to the first bifurcation point $S_e$ a further increase in the pump
intensity, leads to the growth of $\omega_c$ till the second critical point
$S_e$ is achieved. We have analyzed behavior of pulse radiation frequency at
different values of the absorption coefficient $\kappa$. According to
Fig.\ref{fig5} an increase in $\kappa$ at fixed saturation amplitude magnitudes
leads to the shift of minimal and maximal values of $\omega_c$ despite a
topology of the dependence $\omega_c(S_e)$ is not changed. Obtained results are
in good corresponding with experimental observations of such dependence
\cite{casperson}.

Therefore, the dispersion in the relaxation time of the electric field
amplitude $E$, promoting by the absorbing influence, leads to formation of the
stable periodic radiation at saturation amplitude $E_s\simeq 1$.

\subsection{Influence of external modulator}

\begin{figure}[!h]
\centering
\includegraphics[width=70mm]{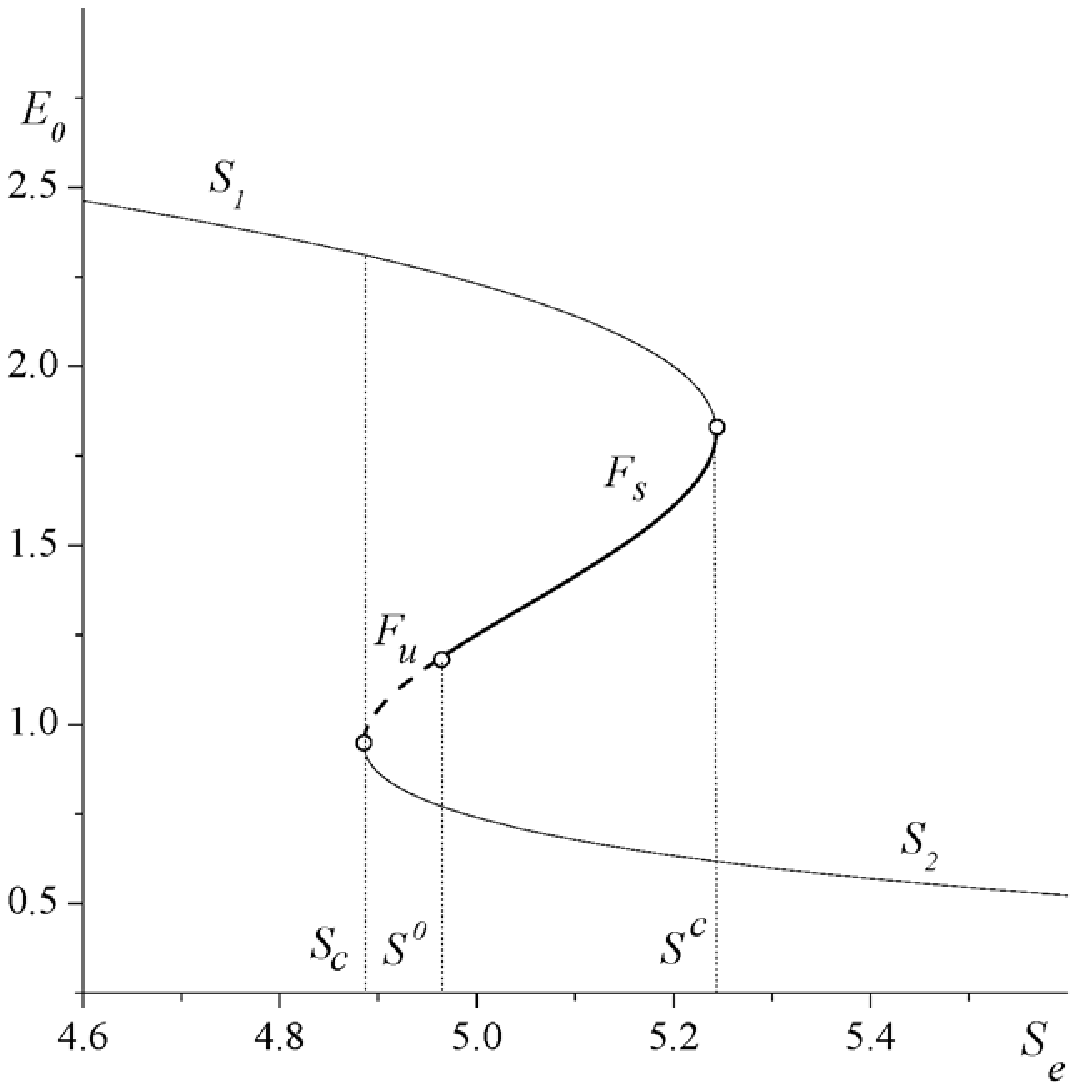}
\caption{Stationary amplitude $E_0$ vs. $S_e$ at $A=1.9$,
$C=-0.455$\label{fig6}}
\end{figure}

\begin{figure}[!h]
\centering
\includegraphics[width=70mm]{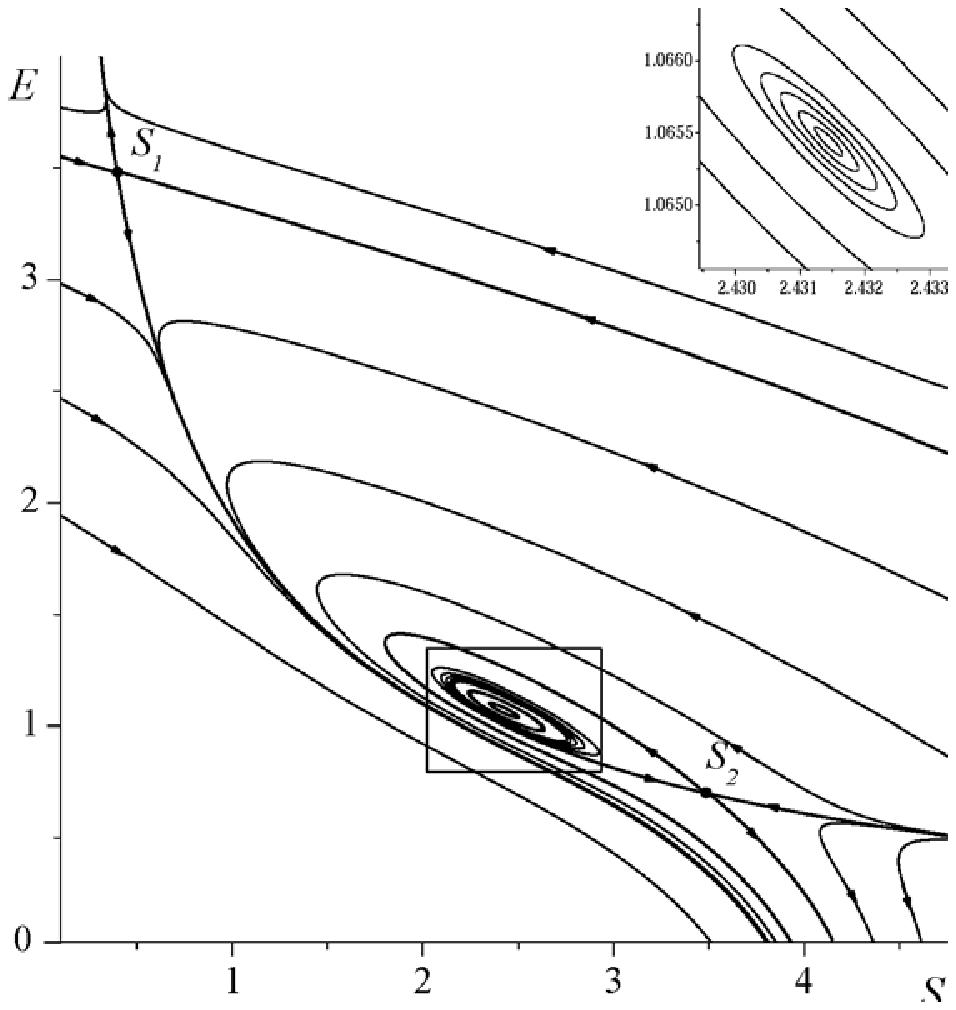}
\caption{Typical phase portrait of the system with external source at
$S_e=5.19152$ $A=1.9$, $C=-0.33025$\label{fig7}}
\end{figure}

Let us consider an influence of the external source $f_e$ at $f_\kappa=0$. It
is principally important that the periodic radiation is possible only if
parameter that controls nonlinear effects $C<0$. Here stationary behavior of
the field $E_0$ versus pump intensity $S_e$ is shown in Fig.\ref{fig6}.
Analysis of the Floquet exponent shows that limit cycles can be formed only if
a stable focus is transformed into an unstable one and vice versa (see
Fig.\ref{fig6}). Here at $S_e<S_c$ and $S_e>S^c$ the phase portrait is
characterized by single saddle point $S_1$ or $S_2$, respectively. In the
domain $S_c<S_e<S^0$ one gets two saddles $S_1$ and $S_2$, divided by an
unstable focus $F_u$. If $S^0<S_e<S^c$, then such saddles are divided by a
stable focus $F_s$. Only if $S_e=S^0$ we will get a trivial situation, where
$\Re\Phi=0$. It means a formation of nested loops of neutral stability
(Fig.\ref{fig7}).
Therefore, external force suppresses processes of dissipative structure
formation.

\subsection{Combined effect of external modulator and nonlinear absorber}

\begin{figure}[!ht]
 \centering
 \includegraphics[width=70mm]{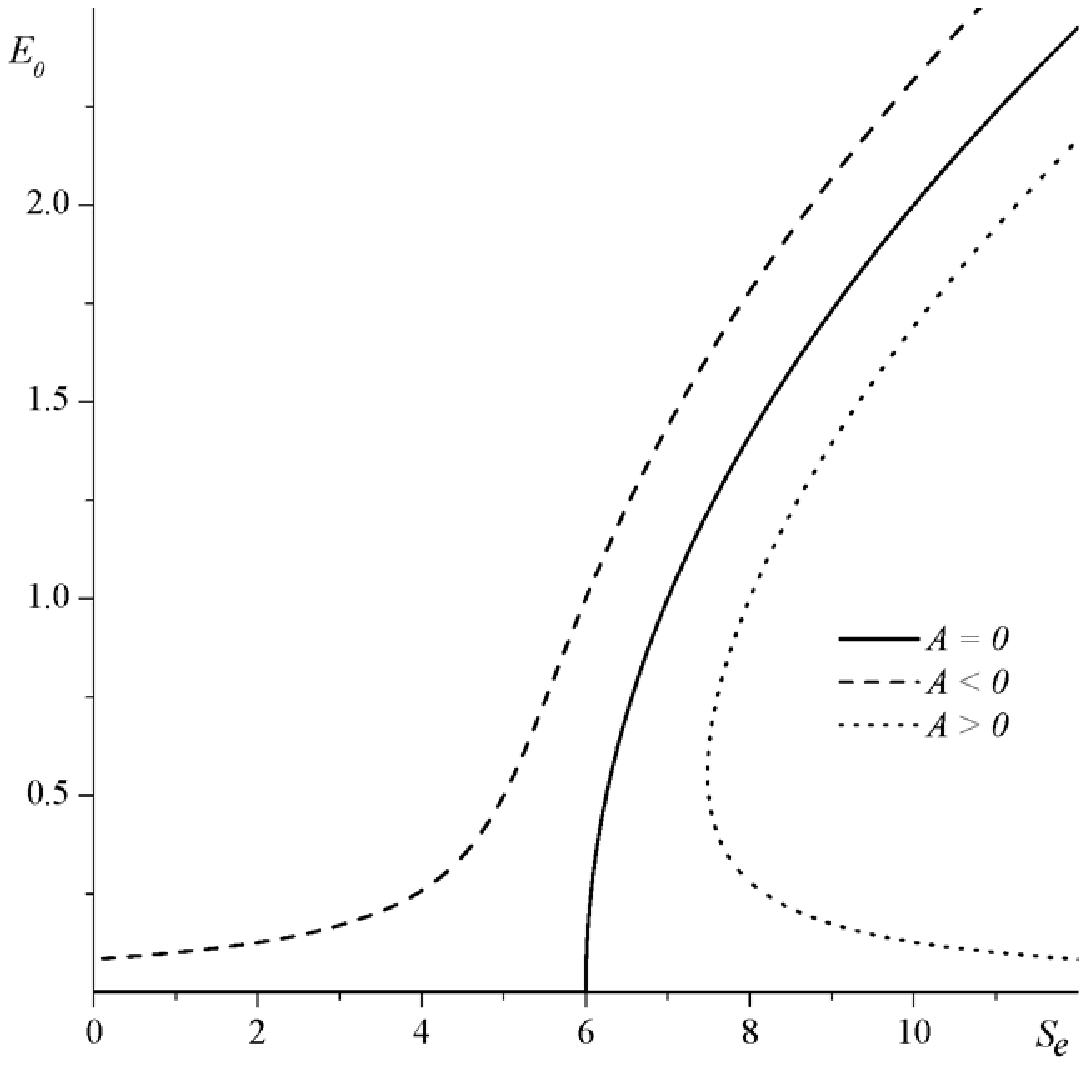}
\caption{Stationary value of electric field amplitude $E_0$ vs. pumping
intensity $S_e$ for the system (\ref{sec4eq1}) at $C=0$, $\kappa=5.0$ and
different magnitudes for the parameter of spontaneous emission $A$\label{fig8}}
\end{figure}

Now we consider an influence of both external modulator and nonlinear absorber
on the processes of dissipative structure formation. Setting $\dot{E}=
\dot{S}=0$ in the system (\ref{sec4eq1}), one gets stationary values of the
electric field amplitude $E_0$ shown in Fig.\ref{fig8}.
As it is seen, if the modulator is turned off ($A=C=0$) then we have a single
stable state with no radiation at small values of the pump parameter $S_e$. If
the threshold given by expression $S_e^c=1+\kappa$ is crossed, then a new
solution of the steady state equation appears and we have a stationary
radiation with an amplitude $E_0\ne0$ which increases with an increase in the
pump intensity. If we set $A<0$ at $C=0$ then we will get a single stable
solution on the whole axis of the pump parameter magnitudes which defines the
radiation amplitude $E_0$. In the opposite case of $A>0$ one gets two
stationary solutions, only if the energy barrier $S_e^c$, given by the solution
of equation $S_e^c = f(A,\kappa_0,\kappa)$, is overcame.

\begin{figure}[!ht]
 \centering
 a) \includegraphics[width=70mm]{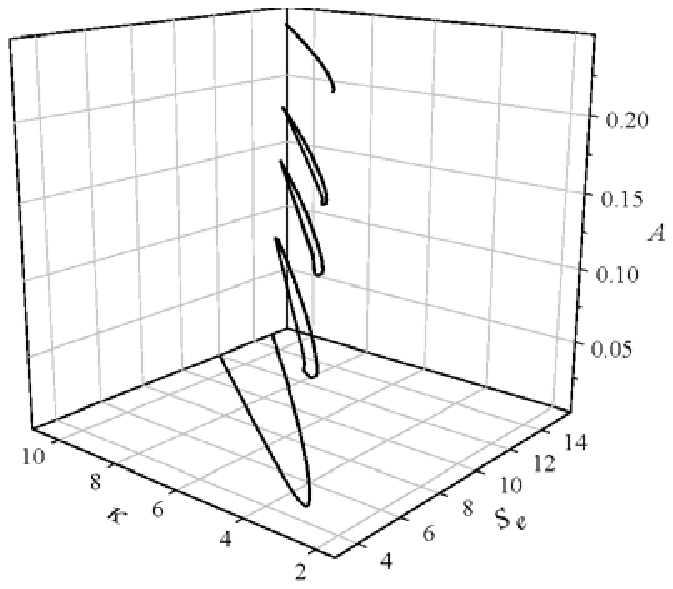}\\
 b) \includegraphics[width=70mm]{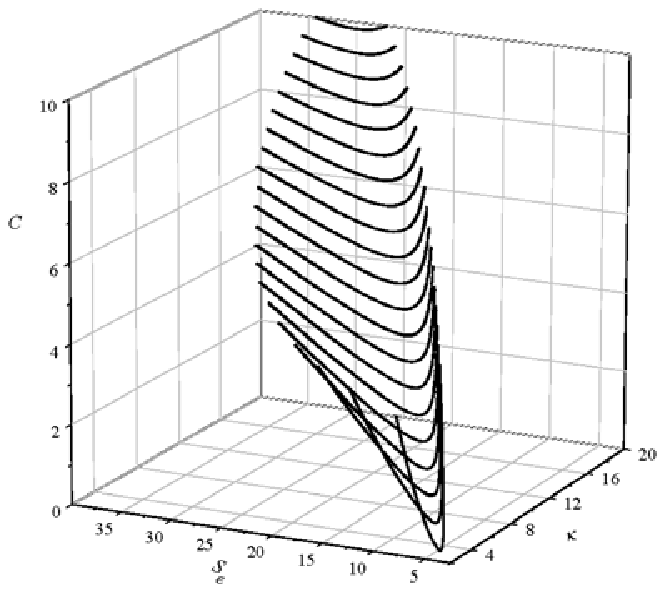}
\caption{Phase diagrams of periodic radiation effect: (a) an influence of the
spontaneous emission intensity $A$ at $C=0$; (b) influence of the photon
scattering intensity $C$ at $A=0$ \label{fig9}}
\end{figure}

Next, we investigate conditions where stable periodic radiation can be
realized. To this end we need to determine a domain defined by conditions
$\lambda=0$ and $\Re\Phi<0$ where periodic solutions of the system
(\ref{sec4eq1}) are exist. Corresponding solutions of the Eq.(\ref{sec4eq28})
are shown in Fig.\ref{fig9}. It illustrates domains of the absorption
coefficient $\kappa$ and pump intensity $S_e$ magnitudes at different
intensities $A$, $C$ where the stable radiation process is realized.
As Fig.\ref{fig9} shows, if we set an absorber inside the cavity only, then a
semi-limited domain of $\kappa$ and $S_e$ magnitudes is formed; inside of this
domain the stable periodic radiation is possible. Introducing a modulator with
$A>0$, $C=0$ (see Fig.\ref{fig9}a), such a domain becomes totally limited.
Moreover, an increase in the parameter $A$ leads to restriction of the values
for the collective parameter $\kappa$ and pump intensity $S_e$, at which one
has stable periodic radiation. At large values $A$ such domain is degenerated
into the line. From Fig.\ref{fig9}b one can see that an increase in the $C$ at
$A=0$ leads to extension of the domain of stable periodic radiation that occurs
at large magnitudes of pump intensity parameter.

\begin{figure}[!ht]
 \centering
 a)\includegraphics[width=70mm]{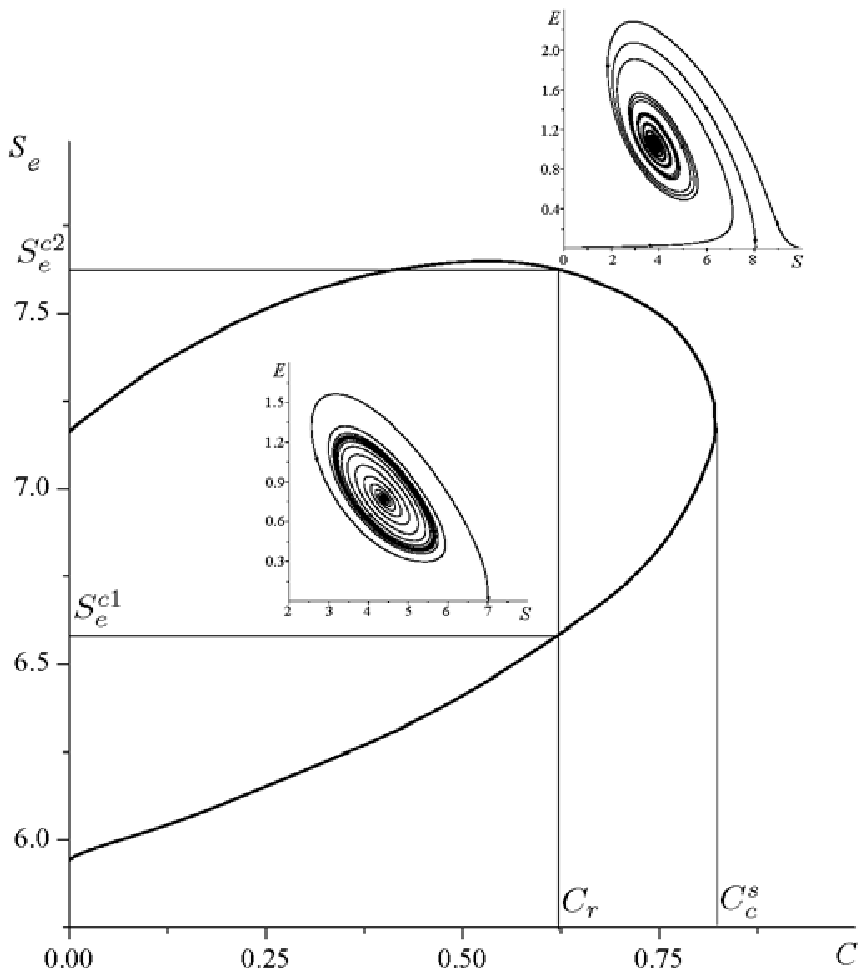}\\
 b)\includegraphics[width=70mm]{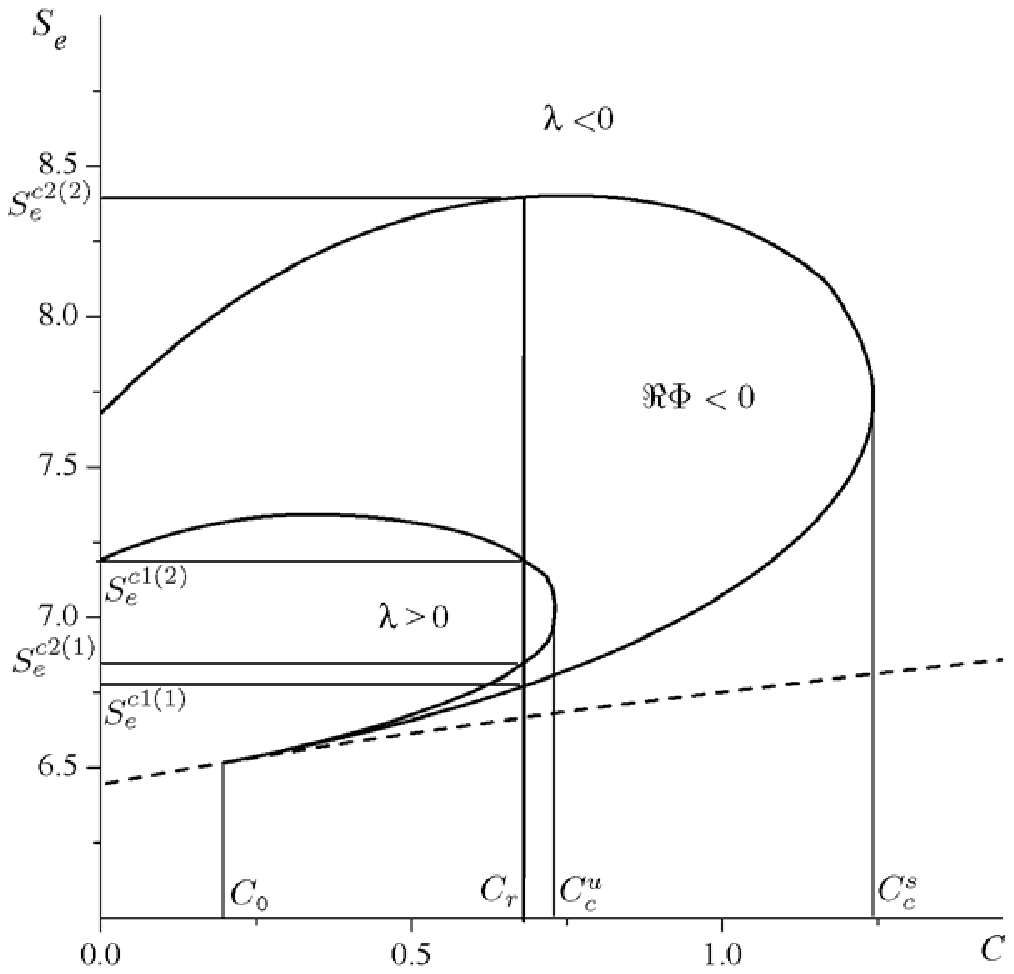}
\caption{ Phase diagrams of Hopf bifurcation: influence of the nonlinear
processes intensity $C$ at $\kappa=5.0$: (a) $A=-0.1$; (b) $A=0.1$
\label{fig10}}
\end{figure}

An influence of nonlinear processes in the modulator on a picture of the stable
periodic radiation formation is presented in Fig.\ref{fig10}.
It is seen, if $A<0$ then there is only stable stationary state (see
Fig.\ref{fig8}) which is a focus ($\Re\Lambda<0$, $\Im\Lambda\ne 0$) on a phase
plane $(E,S)$. Such a fixed point is transformed into a manifold if control
parameters are in the domain including its border shown in Fig.\ref{fig10}a.
Such a manifold is a limit cycle ($\Re\Phi<0$, $\Re\Lambda=0$) in the phase
plane $(E,S)$, that attracts all phase trajectories in the vicinity of it. From
a physical viewpoint it means the formation of the stable pulse periodic
radiation. The domain shown in Fig.\ref{fig10}a is limited by the value of
intensity of nonlinear processes $C_c^s$. One needs to note that if photon
scattering occurs with intensities $C<C_c^s$ then an increase in pump intensity
$S_e$ induces formation of stable periodic radiation at magnitude $S_e^{c1}$
and destroys it at $S_e^{c2}$. In other words, one gets the situation where the
only one reason serves as stimulus for both self-organization and
desorganization.

A picture became more complicated at $A>0$. At first let us discuss the phase
diagram shown in Fig.\ref{fig10}b. At pump limited by the dashed curve in
Fig.\ref{fig10}b there are no stationary solutions and, hence, no stable
regimes of radiation. Next, processes of spontaneous photon annihilation reduce
a domain of stable periodic radiation at pump intensities above dashed curve,
here a domain of unstable behaviour of phase trajectories appears. At small
$C<C_0$ such a stationary regime is defined by the corresponding stationary
solution which is an unstable focus ($\Re\Lambda>0$, $\Im\Lambda\ne 0$).
\begin{figure}[!ht]
 \centering
 a)\includegraphics[width=70mm]{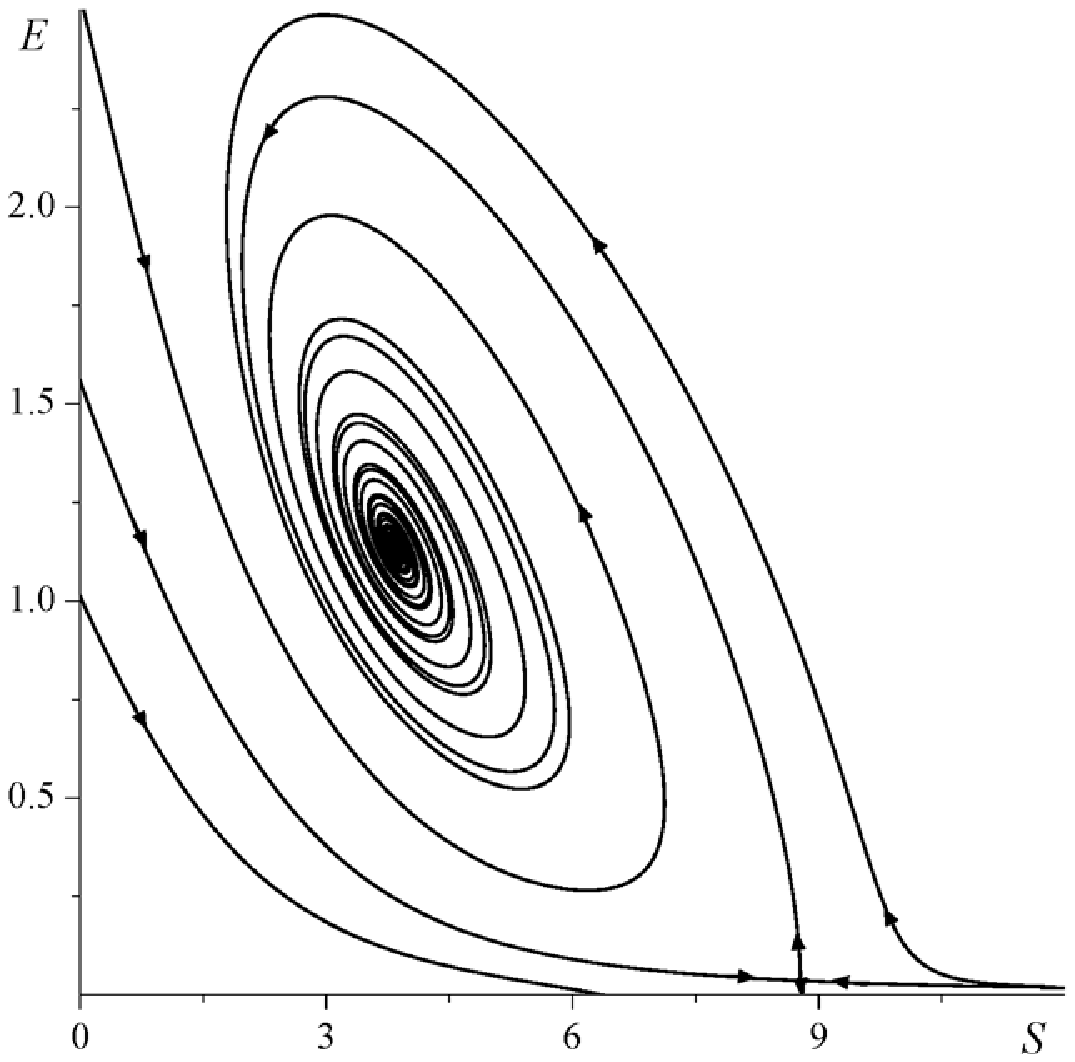}\\ 
 b)\includegraphics[width=70mm]{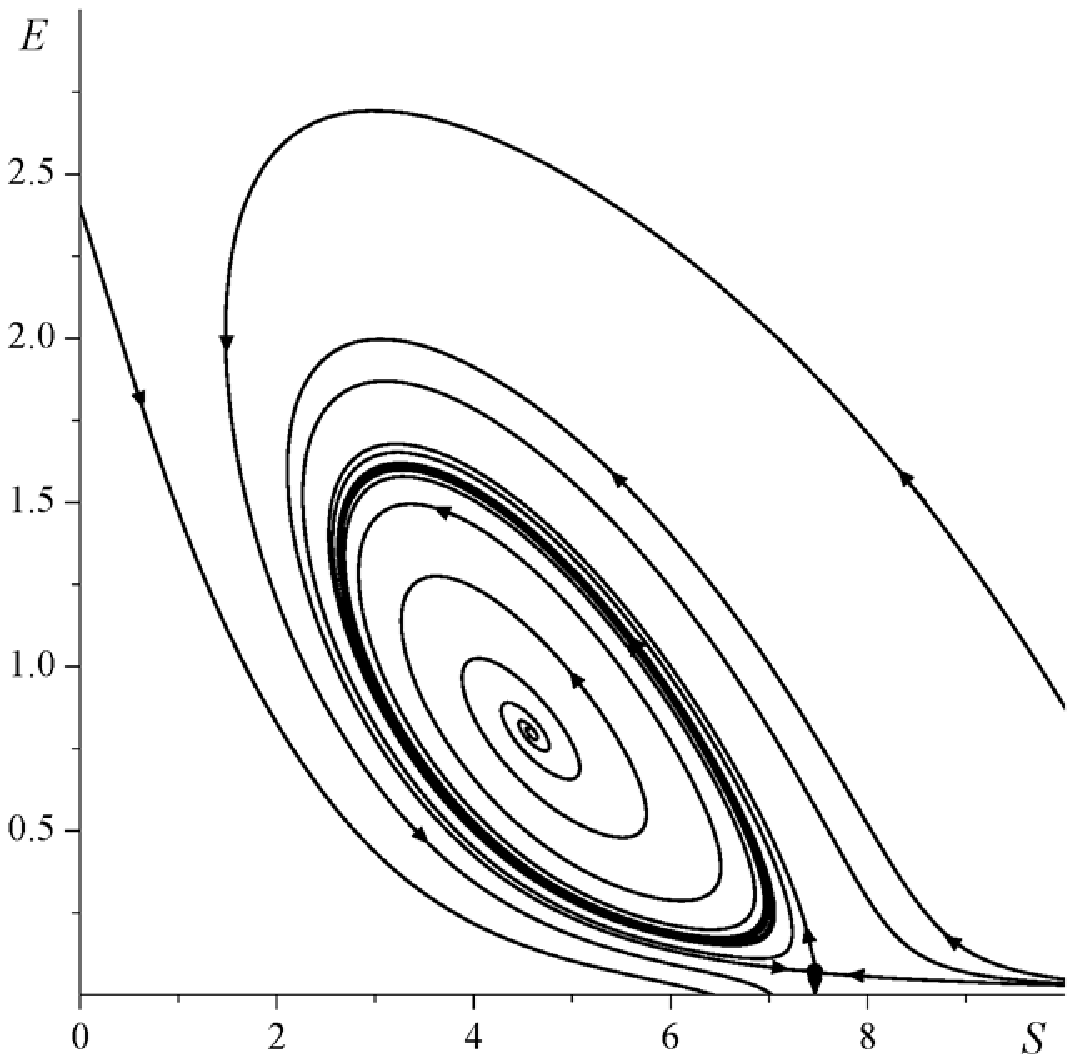}\\ 
 c)\includegraphics[width=70mm]{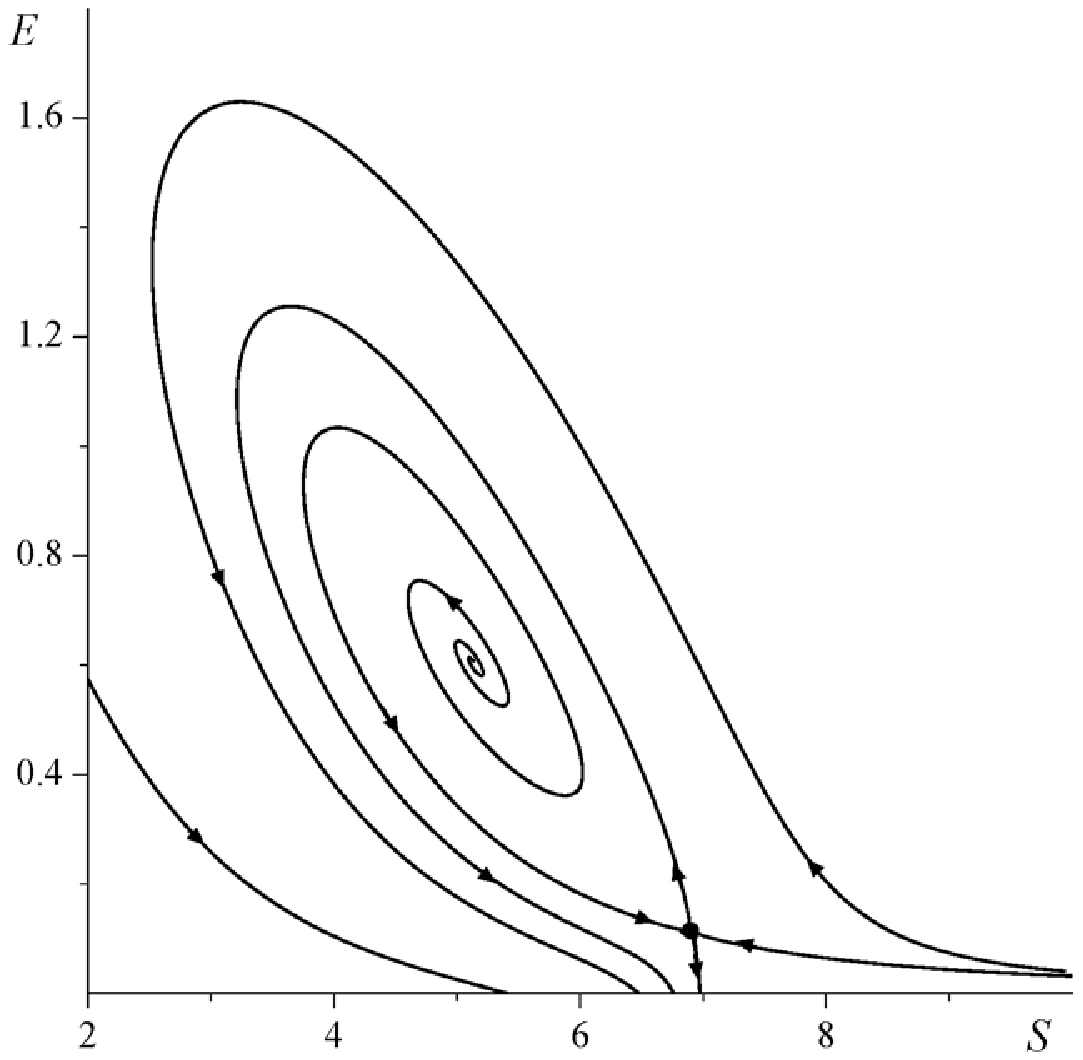}   
\caption{Phase portraits related to the domains in Fig.\ref{fig10}b: (a) --
$A=0.1$, $C=0.5$, $\kappa=5.0$, $S_e=8.8$; (b) -- $A=0.1$, $C=0.5$,
$\kappa=5.0$, $S_e=7.5$; (c) -- $A=0.1$, $C=0.5$, $\kappa=5.0$, $S_e=7.0$}
 \label{fig11}
\end{figure}
The related fixed point is defined as an upper branch of the dashed curve in
Fig.\ref{fig8}. At large values $C_c^u<C<C_c^s$ one has a picture similar to
discussed above. At intermediate values $C_0<C_r<C_c^u$ one can get a very
complicated picture of self-organization. Here with an increase in $S_e$ we
have following picture of transformations: (i) the system passes from the
unstationary regime to stationary one with fixed point to be a stable focus
(Fig.\ref{fig11}a); (ii) at values $S_e^{c1(1)}$ one has a stable periodic
radiation that exists till magnitudes $S_e<S_e^{c2(1)}$ (Fig.\ref{fig11}b);
(iii) a further increase in pump intensity destroys the limit cycle and a
system pass to unstable regime which is characterized by the unstable focus
(Fig.\ref{fig11}c); (iv) if critical value $S_e^{c1(2)}$ is achieved, then a
new Hopf bifurcation occurs and the system evolves according to periodic
trajectories (Fig.\ref{fig11}b); (v) at last, such a coherent regime is
destroyed at $S_e>S_e^{c2(2)}$ (Fig.\ref{fig11}a).

Let us consider more closely properties of phase diagram (Fig.\ref{fig4}),
which shows magnitudes of the absorption coefficient $\kappa$ and the control
parameter $S_e$.
\begin{figure}[!ht]
 \centering
 \includegraphics[width=70mm]{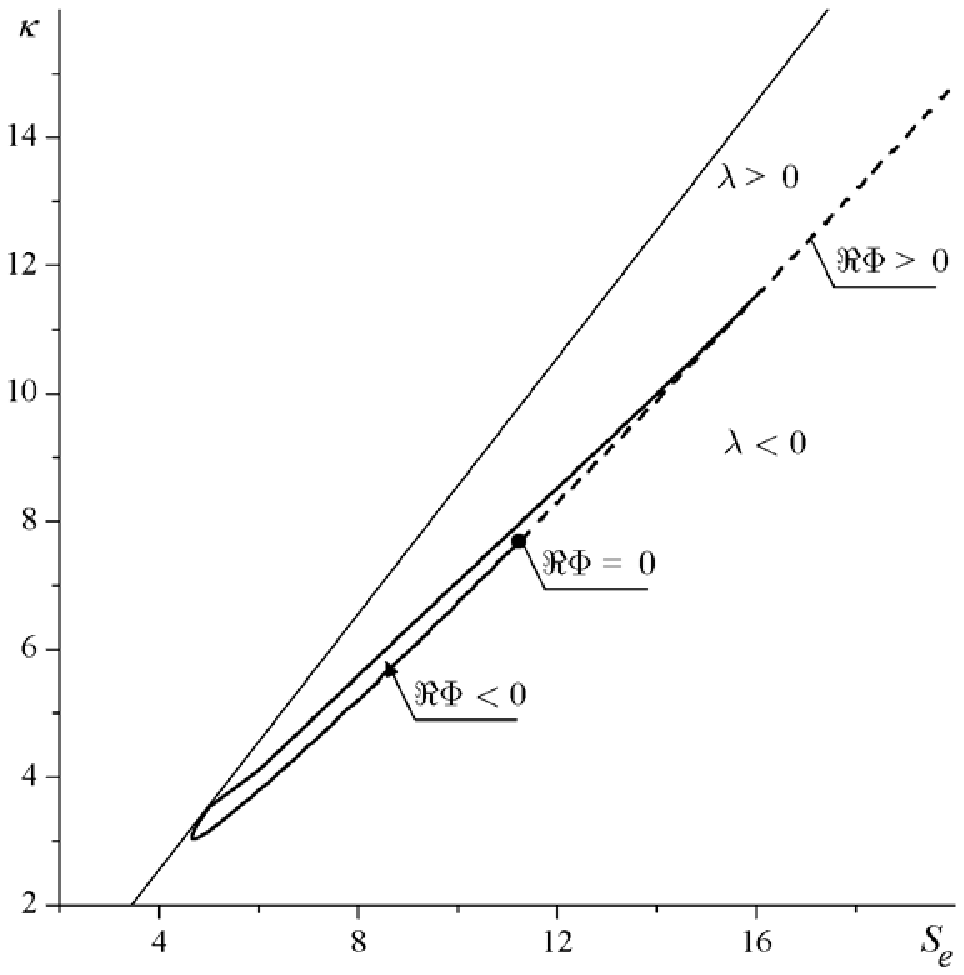}
 \caption{Phase diagram  at $A>0$, $C=0$}
 \label{fig12}
\end{figure}
\begin{figure}[!ht]
 \centering
 a)\includegraphics[width=35mm]{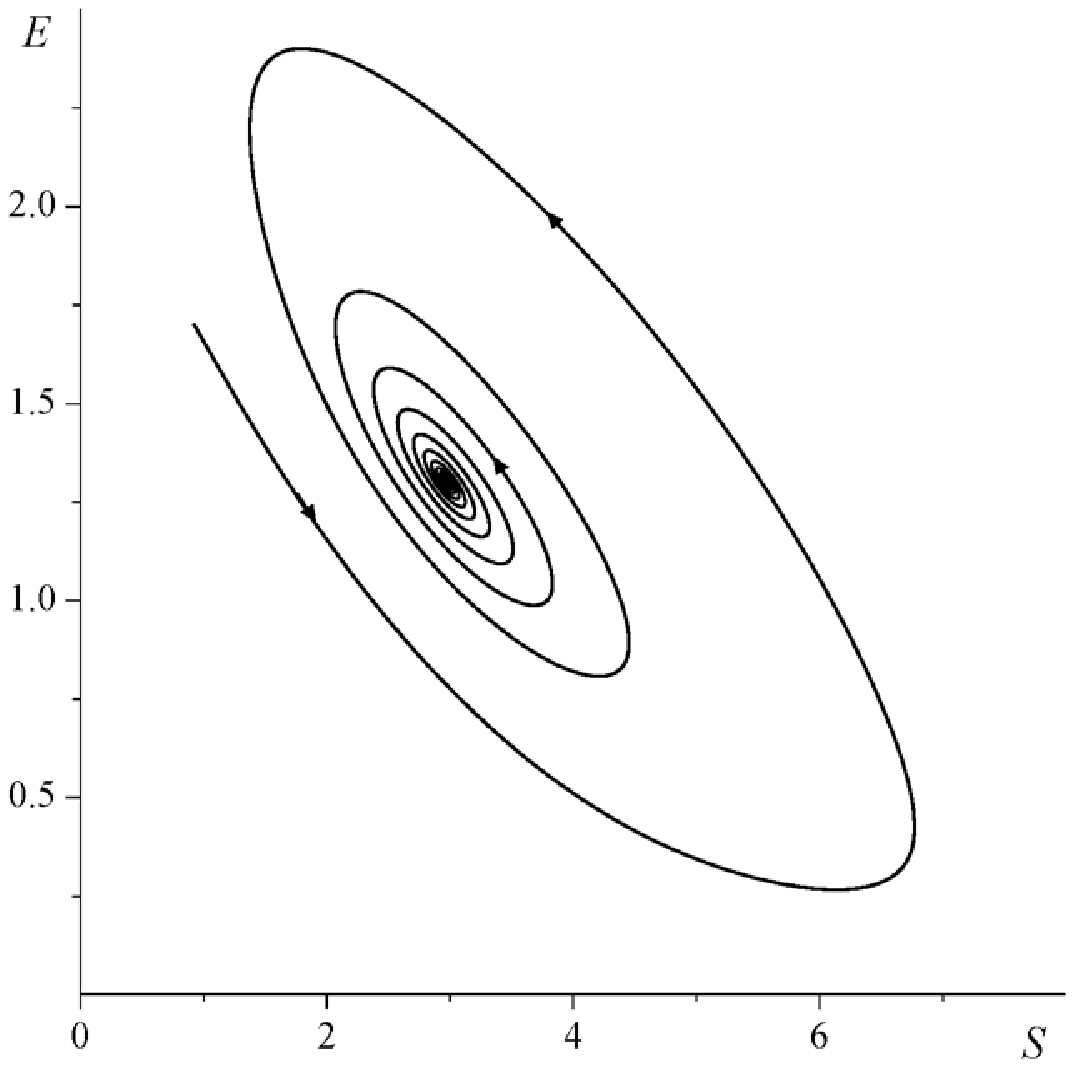} b)\includegraphics[width=35mm]{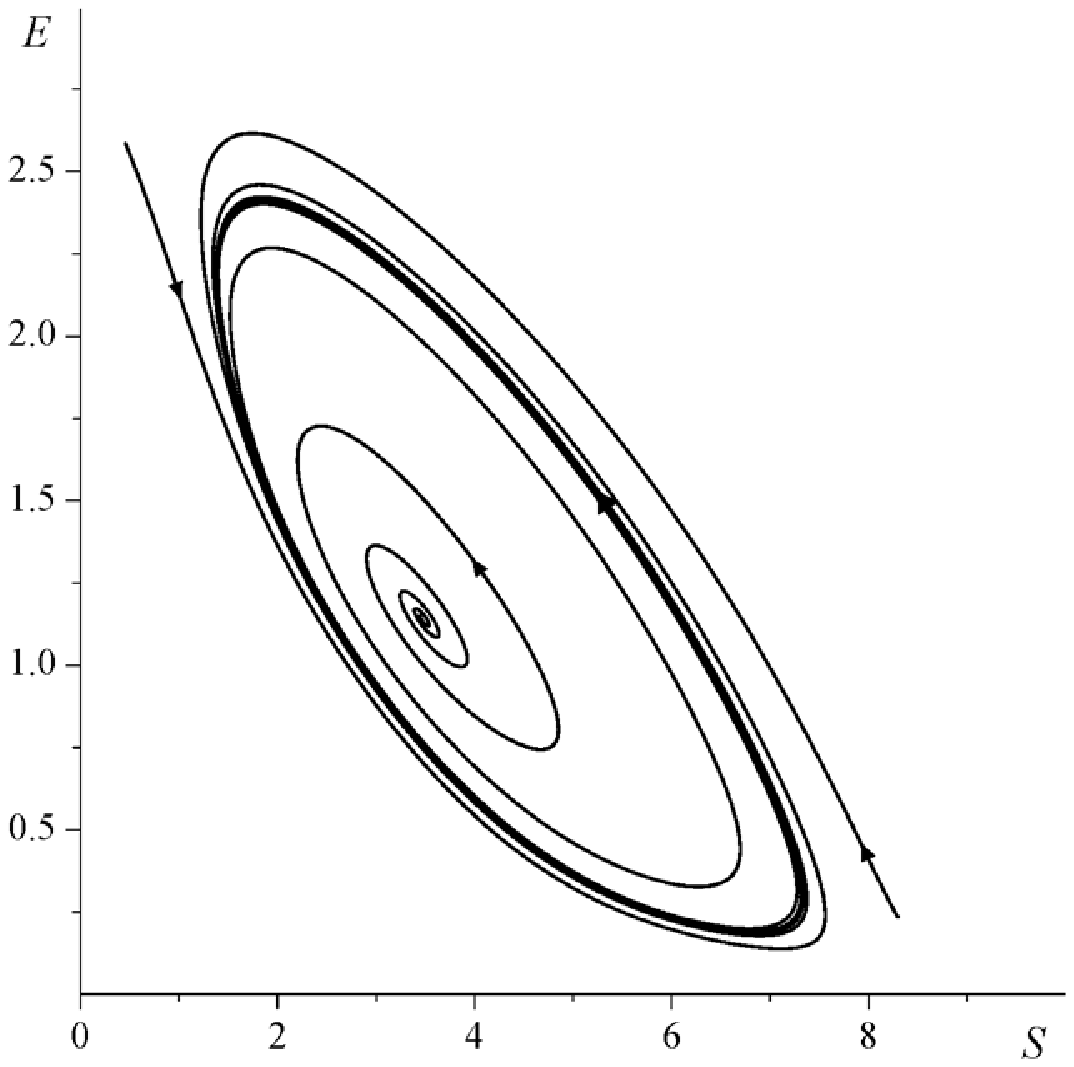}\\
 c)\includegraphics[width=35mm]{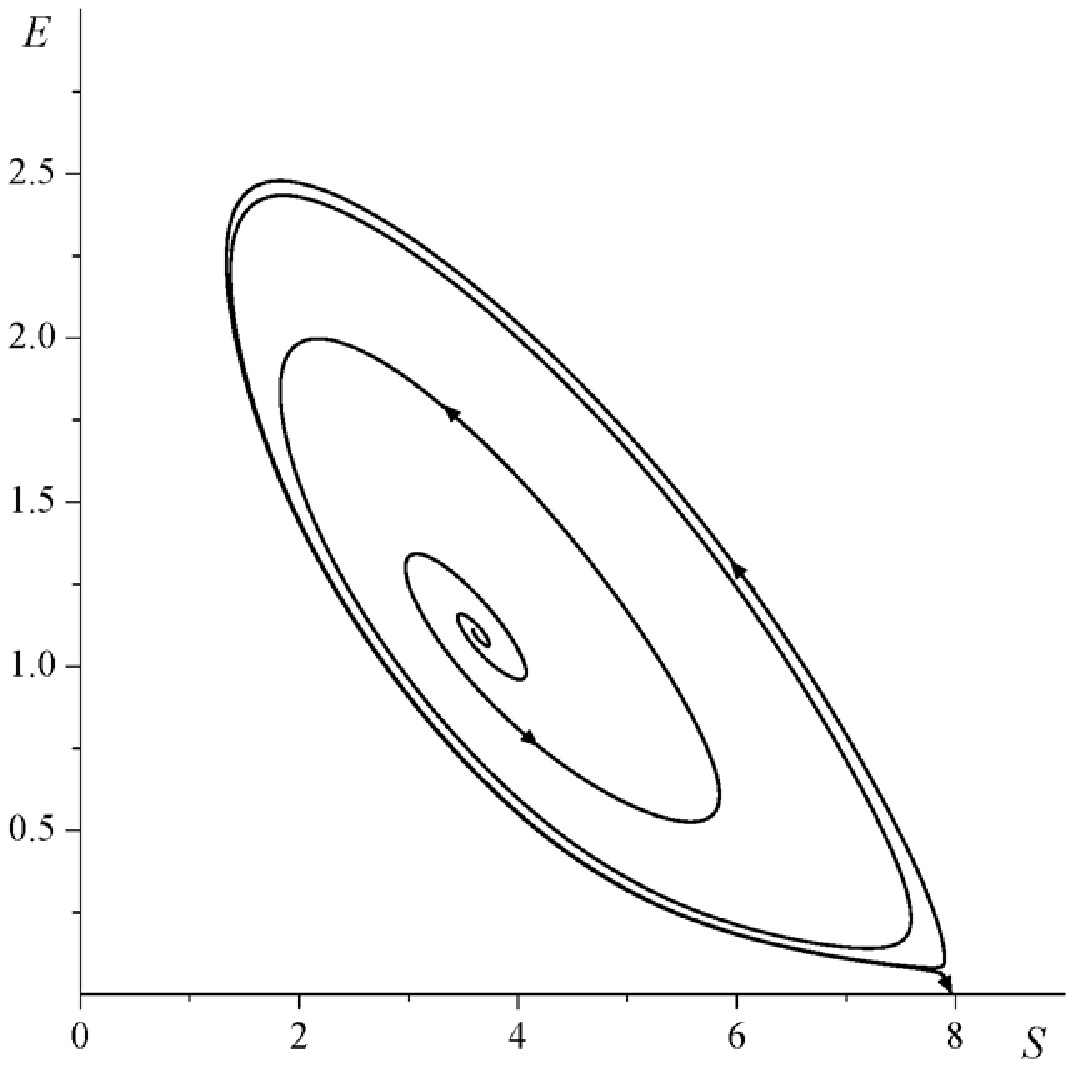} d)\includegraphics[width=35mm]{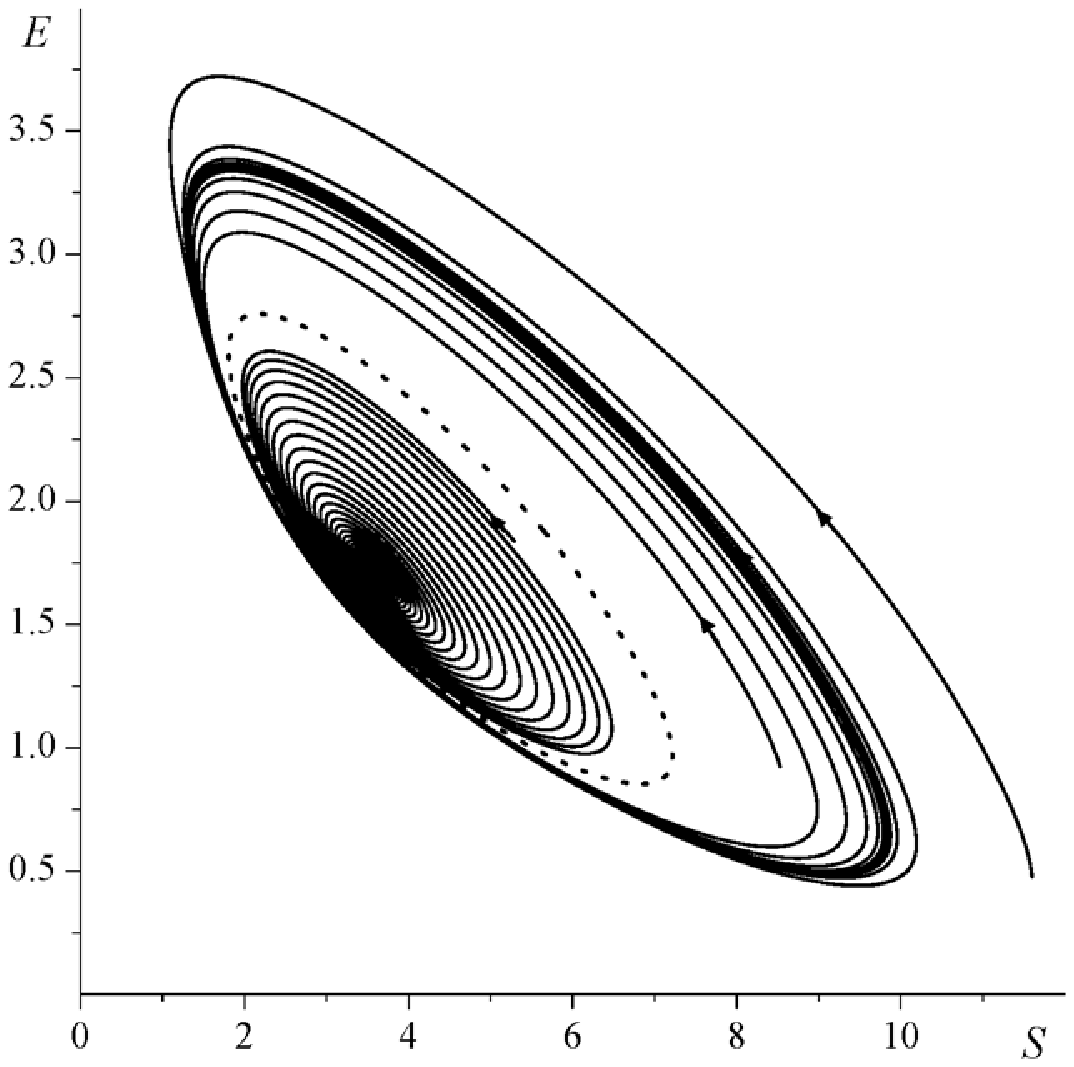}
\caption{Phase portraits related to the phase diagram in Fig.\ref{fig12}: (a)
-- $A=0.1$, $C=0.0$, $\kappa=5.1$, $S_e=8.0$; (b) -- $A=0.1$, $C=0.0$,
$\kappa=5.5$, $S_e=8.0$; (c) -- $A=0.1$, $C=0.0$, $\kappa=5.6$, $S_e=8.0$; (d)
-- $A=0.1$, $C=0.0$, $\kappa=9.86$, $S_e=14.0$.}
 \label{fig13}
\end{figure}
Here the thin solid curve (bifurcation line) defines critical magnitudes for
$\kappa$ and $S_e$ where stationary states appeared. In the domain with
$\Re\Lambda>0$ one has an unstable focus (see Fig.\ref{fig13}a). The stable
limit cycle is realized inside the bounded domain with $\Re\Phi<0$
(Fig.\ref{fig13}b). At small $\kappa$ and large $S_e$ one has a stable focus
(Fig.\ref{fig13}c). Therefore, one gets a transformation of topology of
attractors in the phase plane if parameters $S_e$ or $\kappa$ are changed. An
increase in absorption coefficient $\kappa$ at fixed pump intensity $S_e$ will
produce the oscillating regime (transition from a stable focus to limit cycle).
Such stable periodic regime can be destroyed at large magnitudes of $\kappa$
(transition from the limit cycle into repeller --- unstable focus) and a
further increase in $\kappa$ leads to the absence of any stationary regime at
all. However, the stable periodic solution is observed not on a whole border of
the indicated domain. Figure \ref{fig12} shows that a stable dissipative
structure is formed inside the domain and on the thick solid lines only
($\Re\Phi<0$). A part of the domain border plotted as dashed line corresponds
to conditions $\lambda=0$ and $\Re\Phi>0$, which mean existence of unstable
periodic solution (see Fig.\ref{fig13}d). Hence, there is a point where
$\Re\Phi=0$ and periodic solution changes its stability. In this point the
phase portrait of the system is characterized by a set of nested loops.

\section{Conclusions}

In this Paper we have analyzed properties of self-organization processes in the
two-level class-B laser systems in the presence of absorption effects and
influence of the external force. We have shown that due to the nonlinear
damping the domain of control parameters of the cavity with the stable pulse
radiation is realized. It was shown that varying a saturation amplitude and
absorption coefficient one can pass to different type of radiation,
characterized by fixed point in the phase space type of: stable and unstable
focuses, stable and unstable limit cycles. Introducing the external force that
leads to additional nonlinear effects that reduce domains of control parameters
with stable periodic radiation. It is principally important that due to the
external force influence one can get reentrant Hopf bifurcation. Here there is
a wide range of the external force parameters, where both stable and unstable
dissipative structures are in the phase space. Our results are in good
correspondence with theoretical ones \cite{Khanin1,condmat1} and experimental
observations \cite{Khanin2,casperson,GETF71,PismavGETF,condmat3}.

In our investigation we have considered the simplest case, where relaxation
velocities of the electric field and population inversion are of the same
order. In real systems of the solid-state class-B lasers
$\ve\equiv\varkappa/\gamma_\|\sim10^{-1}\div10^{-3}$, in gas lasers of such
class $\ve\simeq 1$. As was shown theoretically and experimentally
\cite{GETF71} a difference between above relaxation velocities will not change
the picture of stable pulse regime qualitatively. Experimental investigation
shows quantitative changes only.

In our consideration the construction for the external force can be applied to
describe influence of the nonlinear processes: in the nonlinear medium with the
nonlinear dependence of refractive index (a variation of the parameter $C$);
introducing an external incident field with amplitude $A<0$; more complicated
picture with arbitrary $A$ and $C$ under supposition of the dynamic system
stability only.

\end{document}